\documentclass[runningheads]{llncs}
\usepackage[T1]{fontenc}
\usepackage{booktabs}
\usepackage{multirow}
\usepackage{hyperref}
\usepackage{graphicx}
\usepackage{tcolorbox}
\usepackage{graphicx}
\usepackage{arydshln} 
\usepackage[dvipsnames]{xcolor}
\usepackage[table]{xcolor} 
\usepackage{orcidlink}
\usepackage{fontawesome5}
\usepackage{color}

\begin{document}
\title{Cross-Lingual Information Retrieval in Tetun\\for Ad-Hoc Search}
%
%
\author{Altedio Araújo\inst{1}\orcidlink{0009-0001-5417-3981} \and Gabriel de Jesus\inst{2 (\scalebox{0.6}{\faEnvelope[regular]})}\orcidlink{0000-0003-4392-2382} \and Sérgio Nunes\inst{2}\orcidlink{0000-0002-2693-988X}}
\authorrunning{Araújo et al.}
%
%
\institute{School of Management and Technology, Polytechnic of Porto (ESTG), Portugal\\
\email{8220759@estg.ipp.pt}\\
\and INESC TEC / Faculty of Engineering, University of Porto (FEUP), Portugal\\ 
\email{gabriel.jesus@inesctec.pt, sergio.nunes@fe.up.pt}\\
}
\maketitle              
\begin{abstract}
Developing information retrieval (IR) systems that enable access across multiple languages is crucial in multilingual contexts. 
In Timor-Leste, where Tetun, Portuguese, English, and Indonesian are official and working languages, no cross-lingual information retrieval (CLIR) solutions currently exist to support information access across these languages. 
This study addresses that gap by investigating CLIR approaches tailored to the linguistic landscape of Timor-Leste.
Leveraging an existing monolingual Tetun document collection and ad-hoc text retrieval baselines, we explore the feasibility of CLIR for Tetun. 
Queries were manually translated into Portuguese, English, and Indonesian to create a multilingual query set. 
These were then automatically translated back into Tetun using Google Translate and several large language models, and used to retrieve documents in Tetun.
Results show that Google Translate is the most reliable tool for Tetun CLIR overall, and the Hiemstra LM consistently outperforms BM25 and DFR BM25 in cross-lingual retrieval performance.
However, overall effectiveness remains up to 26.95 percentage points lower than that of the monolingual baseline, underscoring the limitations of current translation tools and the challenges of developing an effective CLIR for Tetun. 
Despite these challenges, this work establishes the first CLIR baseline for Tetun ad-hoc text retrieval, providing a foundation for future research in this under-resourced setting.

\keywords{Cross-lingual information retrieval \and Tetun \and Ad-hoc text retrieval \and Low-resource language}
\end{abstract}

\section{Introduction}
Cross-lingual information retrieval (CLIR) systems are essential for enabling users to submit queries in one language and retrieve relevant documents written in another. Such systems are particularly valuable in linguistically diverse contexts, such as Timor-Leste. As a multilingual country, Timor-Leste designates Tetun and Portuguese as its official languages, and English and Indonesian as working languages~\cite{vasconcelos-et-al-2011}, with over 30 dialects spoken across the territory~\cite{de-jesus-2023-ecir}. Within this complex linguistic landscape, CLIR becomes a crucial tool for facilitating access to information across these languages.

A classical CLIR system typically consists of a translation step followed by monolingual retrieval~\cite{zhang-et-al-2019-improving-lrl-clir}. Two primary translation strategies are commonly employed: \textbf{query translation}, where the user’s query is translated into the language of the documents, and \textbf{document translation}, where documents are translated into the language of the query~\cite{ogundepo-et-al-2022,zhou-2012-translation-techniques}. Query translation is generally more efficient and conducive to rapid experimentation, whereas document translation can be computationally expensive and time-intensive, especially when translating an entire corpus~\cite{lin-et-al-2304-01019-simple-effective-neural-ranking-reranking,saleh-pecina-2020-doc-query-translation-clir}.

Tetun, a dialect that became one of the official languages following the country’s restoration of independence from Indonesia in 2002, remains under-resourced, with few algorithms and tools available to support the language. Notably, the first initiative focused on Tetun text information retrieval (IR) began in 2022~\cite{de-jesus-2023-ecir,de-jesus-2022-ir-preliminary}, leading to the creation of foundational resources for IR and natural language processing (NLP)~\cite{de-jesus-nunes-2024-labadain30k-sigul,de-jesus-nunes-labadain30k-dataset-inesctec,de-jesus-nunes-2025-avaliador}, including a test collection and baselines for the ad-hoc text retrieval task~\cite{de-jesus-nunes-2024-baselines}.

Building on these efforts, and given the lack of existing CLIR for Tetun, this study explores the effectiveness of CLIR for retrieving Tetun documents using queries in Portuguese, English, and Indonesian. Results from this study are compared against existing monolingual Tetun ad-hoc text retrieval baselines~\cite{de-jesus-nunes-2024-baselines} to evaluate the performance of CLIR for Tetun. 
This study is guided by the following research questions (RQs):
\begin{itemize}
    \item []\textbf{RQ1:} How does the quality of query translation influence the effectiveness of Tetun ad-hoc text retrieval?
    \item []\textbf{RQ2:} How does the effectiveness of CLIR compare to that of monolingual retrieval in Tetun?
\end{itemize}

To address these RQs, we first manually translated the original Tetun queries from \textit{Labadain-Avaliadór}~\cite{de-jesus-nunes-2025-avaliador} into Portuguese, English, and Indonesian, resulting in a multilingual query set. These translated queries served as input to the CLIR system for retrieving documents in Tetun. To enable retrieval, the multilingual queries were automatically translated back into Tetun using both Google Translate and large language models (LLMs) before being submitted to the Tetun monolingual retrieval system.

To respond to RQ1, we evaluate the quality of the translated queries using the BLEU metric~\cite{acl-papineni-2002-bleu}, with the original Tetun queries serving as reference texts. This RQ aims to examine whether translation quality directly influences the effectiveness of the CLIR system. For RQ2, we compare the retrieval performance of the CLIR system to the established monolingual baselines, evaluating effectiveness using P@10, MAP@10, and NDCG@10 for each source language.

The findings provide initial insights into the feasibility of CLIR for Tetun, demonstrating the impact of translation quality on retrieval effectiveness. This highlights the potential of machine translation-based query methods to enhance the performance of Tetun ad-hoc text retrieval.

The remainder of this paper is organized as follows. Section~\ref{sec:background-related-work} reviews the background and related work. Section~\ref{sec:methodology} presents the methodology, followed by the CLIR experimental setup in Section~\ref{sec:experiments}. Section~\ref{sec:evaluation-results} reports the evaluation and results, which are further discussed in Section~\ref{sec:discussion}. Finally, Section~\ref{sec:conclusion} summarizes the main conclusion and outlines directions for future work.

\section{Background and Related Work} \label{sec:background-related-work}
Tetun is an Austronesian language spoken in Timor-Leste, an island country in Southeast Asia. Historically used as a lingua franca in trade and religious activities, Tetun became one of Timor-Leste's official languages when the country restored its independence in May 2002. It is a low-resource language spoken by approximately 79\% of the country's 1.18 million population, according to the 2015 census~\cite{census-2015-language,de-jesus-nunes-2024-labadain30k-sigul}.
The development of Tetun IR and NLP algorithms and tools began in 2022~\cite{de-jesus-2022-ir-preliminary}, leading to the creation of key resources, including a text dataset~\cite{de-jesus-nunes-labadain30k-dataset-inesctec} and a test collection with baselines for the ad-hoc text retrieval task~\cite{de-jesus-nunes-2024-baselines}. These tools and resources provide the essential foundation for investigating CLIR in Tetun.

CLIR has been extensively studied, with several works exploring its application to low-resource languages (LRLs). Adeyemi et al.~\cite{adeyemi-et-al-2024-ciral} conducted CLIR experiments covering several African languages, including Hausa, Somali, Swahili, and Yoruba, using English queries. Their study investigated both manual and automatic query translation from English into the target languages, and automatic translation of documents into English.
For query translation, they used Google Translate,
and reported that human-translated queries retrieved using BM25 outperformed automatically translated ones by 5.78 percentage points in NDCG@20.
In the CLEF 2009 track~\cite{ferro-peters-clef-2009}, the use of Google Translate for query translation significantly improved cross-lingual retrieval performance. 
Similarly, a study focusing on English to Chinese retrieval showed that translating queries using Google Translate could achieve performance comparable to monolingual baselines, particularly when queries are longer~\cite{wu-2010-query-translation-google-translate}.

Several studies have explored the use of LLMs for query translation in  CLIR. Adeyemi et al.~\cite{adeyemi-et-al-2024-zero-shot-llm} investigated the effectiveness of LLMs for query translation by adopting zero-shot prompting with GPT-3.5 and GPT-4 to translate English queries into Hausa, Somali, Swahili, and Yoruba. The translation quality was evaluated using BLEU~\cite{acl-papineni-2002-bleu}, and GPT-4 achieved the highest scores for all languages.
Similarly, Valentini et al.~\cite{valentini-2025-clirudit} translated both queries and documents from English to French using zero-shot prompting with GPT-4o-mini and LLaMA-3.2, evaluating translation quality using several metrics, including BLEU. For document translation, GPT-4o-mini achieved a higher BLEU score than LLaMA-3.2 and was subsequently used for query translation.
For retrieval, BM25 was used, and document translation with GPT-4o-mini achieved better performance than LLaMA-3.2 in terms of Recall@100 and NDCG@10.

Pecina et al.~\cite{pecina-et-al-2014-mt-mlir-medical-domain} investigated CLIR in the medical domain, focusing on improving query translation quality and examining its impact on retrieval effectiveness. They used Moses~\cite{koehn-et-al-2007-moses}, a phrase-based statistical machine translation system trained on medical domain data, and compared its performance with Google Translate and Microsoft Bing Translator. Retrieval was conducted in English using the CLEF eHealth 2013 collection, while the original queries (written in Czech, German, and French) were translated into English. 
In terms of translation quality, Moses outperformed both Google Translate and Microsoft Bing Translator. For retrieval, Moses maintained superior performance compared to Bing Translator in Czech to English translation at P@10, but it did not consistently outperform Google Translate across all input languages. Overall, the authors reported that higher BLEU scores in translation did not lead to improved retrieval effectiveness.

Likewise, Lignos et al.~\cite{lignos-et-al-challenge-mt-lrl} examined the relationship between machine translation quality and retrieval effectiveness in CLIR for Czech and German, focusing on document translation. In their study, queries were written in English, and documents in Czech and German were translated into English. In line with the findings of Pecina et al.~\cite{pecina-et-al-2014-mt-mlir-medical-domain}, higher BLEU scores in document translation did not enhance retrieval performance.

Building on this foundation, and given recent advancements in LLMs and the inclusion of Tetun to Google Translate,\footnote{\url{https://blog.google/products/translate/google-translate-new-languages-2024/}} new opportunities have emerged to explore the applicability of these approaches to Tetun. In this study, we evaluate the application of Google Translate alongside several LLMs to assess their effectiveness in translating queries into Tetun and compare their performance against established monolingual Tetun IR baselines.

\section{Methodology} \label{sec:methodology}
To investigate CLIR for Tetun, we use \textit{Labadain-Avaliadór}~\cite{de-jesus-nunes-2025-avaliador}, a Tetun test collection for the ad-hoc text retrieval task, which contains 33,500 documents, 59 queries, and 5,900 relevance judgments (qrels)~\cite{de-jesus-nunes-2024-baselines}.
The study began by manually translating Tetun queries into Portuguese, English, and Indonesian. The translations were conducted by three native Tetun-speaking volunteer students, each responsible for a single target language in which they are fluent. The process followed established guidelines, including specific instructions to preserve abbreviations and named entities, resulting in a multilingual query set.

These multilingual queries are subsequently used as input for retrieval. During the query processing phase, the input queries are automatically translated back into Tetun using Google Translate and LLMs, including GPT-4.5, DeepSeek-Chat, Claude 3.7 Sonnet, and LLaMA 3.3 70B. To assess translation quality, the outputs are evaluated using the BLEU metric~\cite{acl-papineni-2002-bleu}, with the original Tetun queries serving as reference texts.

The translated queries are then used for retrieval. To enable performance comparison, we adopt the best-performing retrieval strategy previously reported in monolingual Tetun ad-hoc text retrieval. The retrieval and ranking models used in the experiments are BM25~\cite{robertson-zaragoza-2009}, DFR BM25~\cite{amati-rijsbergen-2002}, and the Hiemstra language model (LM)~\cite{robertson-hiemstra-2001}. These models are evaluated using Precision, MAP, and NDCG at rank ten ($k=10$) for each input query in English, Portuguese, and Indonesian.
Results from CLIR are compared to the monolingual Tetun ad-hoc text retrieval baselines~\cite{de-jesus-nunes-2024-baselines} to assess their relative performance.

\section{Experimental Setup} \label{sec:experiments}
The experiments are organized into two parts: query translation and cross-lingual retrieval. Three query translation strategies were evaluated: Google Translate, LLMs with zero-shot prompting, and LLMs with few-shot prompting using one example. The experimental setup for each configuration is described below.

\subsection{Query Translations}\label{subsec:query-translations}
Query translation using Google Translate follows the official documentation and default settings of the basic translation endpoint\footnote{\url{https://cloud.google.com/translate/docs/reference/rest/v2/translate}} provided by the Google Translate API. For LLM-based translations, the same prompt is used across all models, with the temperature set to zero to ensure deterministic outputs.

The prompt used to instruct the LLMs was compiled based on best practices for prompting translation models~\cite{adeyemi-et-al-2024-zero-shot-llm,sui-he-prompting-gpt-translation-2024,tang-et-al-2025-adaptive-few-shot-prompting-for-machine-translation}.\footnote{\url{https://docs.anthropic.com/en/resources/prompt-library/polyglot-superpowers}} Details of the zero-shot prompting are presented in Prompt~\ref{zero-shot-prompt}. 

\begin{tcolorbox}[
colback=gray!10, 
colframe=black, 
title=Prompt 4.1. Zero-shot prompt used for query translation with LLMs,
boxrule=0.4pt, 
label=zero-shot-prompt
]
You are an expert translator. Your task is to translate the following user search query from \textit{\{source\_lang\}} into Tetun.\\

Query: \textit{\{query\}}\\

Translate the query accurately, using the official grammar of Tetun and correct vocabulary. Return only the translation and do not include any additional text.

\end{tcolorbox}

For few-shot prompting, we used the same base prompt as in the zero-shot setting, with one example query and its corresponding translation in each target language included in the prompt for each LLM call. The additional instruction appended to the base prompt is shown below:

\begin{tcolorbox}[
    colback=gray!10,
    boxrule=0pt,
    colframe=white,
    sharp corners,
    left=1mm,
    right=1mm,
    top=0mm,
    bottom=0mm
]
    Refer to the example below to guide your translation:\\

    \textit{\{source\_lang\}} query: \textit{\{example\_query\}}\\
    Tetun translation: \textit{\{example\_translation\}}
\end{tcolorbox}

\subsection{Cross-Lingual Retrieval}
For retrieval, we employed the monolingual Tetun ad-hoc text retrieval framework adopted by de Jesus and Nunes~\cite{de-jesus-nunes-2024-baselines}. We used PyTerrier~\cite{macdonald-tonellotto-2020}, a Python API for the Terrier IR platform~\cite{ounis-et-al-2005}, for indexing, retrieval, and ranking, using the default settings for each retrieval model.
This setup indexes only the document titles and applies uniform preprocessing to both queries and titles, including lowercasing and the removal of punctuation, special characters, apostrophes, and hyphens. All queries translated from Portuguese, English, and Indonesian into Tetun, using each of the translation techniques described in subsection~\ref{subsec:query-translations}, were employed for retrieval and evaluated using \textit{Labadain-Avaliadór}~\cite{de-jesus-nunes-2025-avaliador}.

\section{Evaluation and Results} \label{sec:evaluation-results}
The evaluation focuses on the quality of query translations and the effectiveness of CLIR compared to monolingual Tetun ad-hoc text retrieval. Translation quality was measured using BLEU~\cite{acl-papineni-2002-bleu}, while retrieval effectiveness was assessed at P@10, MAP@10, and NDCG@10. The evaluation results are shown in Table~\ref{tab:bleu-scores}.

\begin{table}[!htbp]
\centering
\caption{BLEU scores for translation quality from Portuguese (Pt), English (En), and Indonesian (Id) into Tetun (Tdt).}
\label{tab:bleu-scores}
\setlength{\tabcolsep}{6pt}
\begin{tabular}{@{}lrrrrrr@{}}
\toprule
\multicolumn{1}{c}{\multirow{2}{*}{\textbf{Model}}} & \multicolumn{3}{c}{\textbf{Zero-Shot}}              & \multicolumn{3}{c}{\textbf{Few-Shot}}               \\ \cmidrule{2-7} 
\multicolumn{1}{c}{}                                & \textbf{PT-Tdt} & \textbf{En-Tdt} & \textbf{Id-Tdt} & \textbf{Pt-Tdt} & \textbf{En-Tdt} & \textbf{Id-Tdt} \\ \hline
Google Translate                                    & \textbf{\cellcolor{Gray!15}0.4659}          & \cellcolor{Gray!15}0.4530          & \cellcolor{Gray!15}0.5636          & -         & -          & -        \\
GPT-4.5-preview                                     & \cellcolor{Gray!15}0.4156          & \textbf{\cellcolor{Gray!15}0.4882}          & \cellcolor{Gray!15}0.6038          & 0.4215          & 0.4156          & 0.4616          \\
DeepSeek-Chat                                       & \cellcolor{Gray!15}0.3655          & \cellcolor{Gray!15}0.3762          & \cellcolor{Gray!15}0.5091          & 0.4634          & 0.4183          & 0.4994          \\
Claude-3.7 Sonnet                                   & \cellcolor{Gray!15}0.4602          & \cellcolor{Gray!15}0.4649          & \textbf{\cellcolor{Gray!15}0.6256}          & \textbf{0.5378}          & \textbf{0.5412}          & \textbf{0.6425}         \\
LLaMA-3.3 70B                                       & \cellcolor{Gray!15}0.3566          & \cellcolor{Gray!15}0.3184          & \cellcolor{Gray!15}0.4172          & 0.3704          & 0.3122          & 0.4033          \\ \hline
\end{tabular}
\end{table}

As observed, under the zero-shot setting, BLEU scores varied across source languages and models. Google Translate achieved the highest score for Portuguese-to-Tetun translation, GPT-4.5 performed best for English-to-Tetun, and Claude 3.7 Sonnet achieved the highest score for Indonesian-to-Tetun. In the few-shot setting, Claude 3.7 Sonnet recorded the highest BLEU scores for all languages.

When comparing BLEU scores from few-shot and zero-shot LLM-based translations, the following results are observed:
(1) \textbf{Portuguese-to-Tetun:} translation quality improved across all models in the few-shot setting, with gains of up to +9.79 percentage points;
(2)\textbf{ English-to-Tetun:} results varied by model, performance dropped for GPT-4.5-preview and LLaMA-3.3 70B by up to -7.25 percentage points, while DeepSeek-Chat and Claude 3.7 Sonnet showed improvements of up to +7.63 percentage points; and
(3) \textbf{Indonesian-to-Tetun:} only Claude 3.7 Sonnet showed an improvement (+1.69 percentage points), while the other models experienced performance drops of up to -14.22 percentage points.

Regarding the evaluation of retrieval performance, results using zero-shot query translations are presented in Table~\ref{tab:clir-tetun-zero-shot}. The findings show that translating input queries with Google Translate, followed by retrieval and ranking using the Hiemstra LM, generally outperformed the LLM-based approaches. An exception was observed in Portuguese-to-Tetun translation, where LLMs achieved higher scores at P@10 and MAP@10. Nevertheless, all CLIR results remained significantly below the performance of the monolingual Tetun IR baselines, by up to -26.95 percentage points for Portuguese-to-Tetun translation, -25.93 for English-to-Tetun, and -23.22 for Indonesian-to-Tetun.

\begin{table}[!ht]
\centering
\caption{Effectiveness of CLIR using zero-shot translated queries
compared to the monolingual baseline. 
Values in \textbf{bold} indicate the best CLIR performance for each source language. The baseline, highlighted in \colorbox{orange!15}{orange}, indicates the overall best performance across all retrieval metrics.}\label{tab:clir-tetun-zero-shot}
\resizebox{\textwidth}{!}{
\setlength{\tabcolsep}{8pt}
\begin{tabular}{lllrrr}
\hline
\textbf{Translation}                     & \textbf{Language}                     & {\textbf{Model}} & \textbf{P@10} & \textbf{MAP@10} & \textbf{NDCG@10} \\ \hline
Baseline & Tetun 
                                   & DFR BM25                               & \cellcolor{orange!15}0.8390          & \cellcolor{orange!15}0.2804                & \cellcolor{orange!15}0.7356                 
                                   \\ \hline
\multirow{9}{*}{Google Translate}  & \multirow{3}{*}{Portuguese - Tetun} & BM25                                    & 0.5559         & 0.1659                & 0.4934                 \\
                                   &                                       & DFR BM25                               & 0.5576        & 0.1661                & 0.4931                 \\
                                   &                                       & Hiemstra LM                            & 0.5576         & 0.1706                & \textbf{0.4914}                 \\ \cdashline{2-6}
                                   & \multirow{3}{*}{English - Tetun}      & BM25                                    & 0.5712         & 0.1633                & 0.4937                 \\
                                   &                                       & DFR BM25                               & 0.5695         & 0.1634                & 0.4907                 \\
                                   &                                       & Hiemstra LM                            & \textbf{0.5797}         & \textbf{0.1694}                & \textbf{0.4971}                 \\ \cdashline{2-6}
                                   & \multirow{3}{*}{Indonesian - Tetun}   & BM25                                    & 0.6000            & 0.1861                & 0.5216                 \\
                                   &                                       & DFR BM25                               & 0.6017         & 0.1865                & 0.5226                 \\
                                   &                                       & Hiemstra LM                            & \textbf{0.6068}         & \textbf{0.1893}                & \textbf{0.5237}                 \\ \hline
\multirow{9}{*}{GPT-4.5-preview}           & \multirow{3}{*}{Portuguese - Tetun}   & BM25                                    & 0.4864         & 0.1472                & 0.4230                  \\
                                   &                                       & DFR BM25                               & 0.4881         & 0.1474                & 0.4235                 \\
                                   &                                       & Hiemstra LM                            & 0.4983         & 0.1565                & 0.4257                 \\ \cdashline{2-6}
                                   & \multirow{3}{*}{English - Tetun}      & BM25                                    & 0.4966         & 0.1463                & 0.4161                 \\
                                   &                                       & DFR BM25                               & 0.4966         & 0.1467                & 0.4155                 \\
                                   &                                       & Hiemstra LM                            & 0.4847         & 0.1497                & 0.4149                 \\ \cdashline{2-6}
                                   & \multirow{3}{*}{Indonesian - Tetun}   & BM25                                    & 0.5407         & 0.1646                & 0.4576                 \\
                                   &                                       & DFR BM25                               & 0.5424         & 0.1654                & 0.4582                 \\
                                   &                                       & Hiemstra LM                            & 0.5339         & 0.1614                & 0.4495                 \\ \hline
\multirow{9}{*}{DeepSeek-Chat}     & \multirow{3}{*}{Portuguese - Tetun}   & BM25                                    & 0.4407         & 0.1246                & 0.3739                 \\
                                   &                                       & DFR BM25                               & 0.4407         & 0.1247                & 0.3734                 \\
                                   &                                       & Hiemstra LM                            & 0.4441         & 0.1262                & 0.3754                 \\ \cdashline{2-6}
                                   & \multirow{3}{*}{English - Tetun}      & BM25                                    & 0.4373         & 0.1329                & 0.3695                 \\
                                   &                                       & DFR BM25                               & 0.4373         & 0.1337                & 0.3686                 \\
                                   &                                       & Hiemstra LM                            & 0.4407         & 0.1364                & 0.3719                 \\ \cdashline{2-6}
                                   & \multirow{3}{*}{Indonesian - Tetun}   & BM25                                    & 0.5000            & 0.1490                 & 0.4232                 \\
                                   &                                       & DFR BM25                               & 0.5000            & 0.1491                & 0.4222                 \\
                                   &                                       & Hiemstra LM                            & 0.4932         & 0.1439                & 0.4139                 \\ \hline
\multirow{9}{*}{Claude-3.7 Sonnet} & \multirow{3}{*}{Portuguese - Tetun}   & BM25                                    & 0.5475         & 0.1678                & 0.4763                 \\
                                   &                                       & DFR BM25                               & 0.5508         & 0.1689                & 0.4774                 \\
                                   &                                       & Hiemstra LM                            & \textbf{0.5695}         & \textbf{0.1779}                & 0.4912                 \\ \cdashline{2-6}
                                   & \multirow{3}{*}{English - Tetun}      & BM25                                    & 0.5678         & 0.1571                & 0.4656                 \\
                                   &                                       & DFR BM25                               & 0.5627         & 0.1563                & 0.4624                 \\
                                   &                                       & Hiemstra LM                            & 0.5610        & 0.1625                & 0.4717                 \\ \cdashline{2-6}
                                   & \multirow{3}{*}{Indonesian - Tetun}   & BM25                                    & 0.5763         & 0.1868                & 0.4910                  \\
                                   &                                       & DFR BM25                               & 0.5797         & 0.1879                & 0.4920                  \\
                                   &                                       & Hiemstra LM                            & 0.5763         & 0.1833                & 0.4890                  \\ \hline
\multirow{9}{*}{LLaMA-3.3 70B}     & \multirow{3}{*}{Portuguese - Tetun}   & BM25                                    & 0.3831         & 0.1060                 & 0.3197                 \\
                                   &                                       & DFR BM25                               & 0.3814         & 0.1060               & 0.3185                 \\
                                   &                                       & Hiemstra LM                            & 0.3729         & 0.0990                 & 0.3014                 \\ \cdashline{2-6}
                                   & \multirow{3}{*}{English - Tetun}      & BM25                                    & 0.3508         & 0.0896                & 0.2896                 \\
                                   &                                       & DFR BM25                               & 0.3508         & 0.0899                & 0.2887                 \\
                                   &                                       & Hiemstra LM                            & 0.3525         & 0.0921                & 0.2936                 \\ \cdashline{2-6}
                                   & \multirow{3}{*}{Indonesian - Tetun}   & BM25                                    & 0.3763         & 0.1149                & 0.3045                 \\
                                   &                                       & DFR BM25                               & 0.3763         & 0.1145                & 0.3040                  \\
                                   &                                       & Hiemstra LM                            & 0.3780          & 0.1139                & 0.3043 \\ \hline 
\end{tabular}
}
\end{table}

\begin{table}[!ht]
\caption{Effectiveness of CLIR using few-shot translated queries
compared to the monolingual baseline. 
Values in \textbf{bold} indicate the best CLIR performance for each source language. The baseline, highlighted in \colorbox{orange!15}{orange}, indicates the overall best performance across all retrieval metrics.}\label{tab:clir-tetun-few-shot}
\resizebox{\textwidth}{!}{
\setlength{\tabcolsep}{8pt}
\begin{tabular}{lllrrr}
\hline
\textbf{Translation}                     & \textbf{Language}                     & {\textbf{Model}} & \textbf{P@10} & \textbf{MAP@10} & \textbf{NDCG@10} \\ \hline
Baseline & Tetun & DFR BM25                               & \cellcolor{orange!15}0.8390          & \cellcolor{orange!15}0.2804                & \cellcolor{orange!15}0.7356 \\ \hline
\multirow{9}{*}{GPT-4.5-preview}   & \multirow{3}{*}{Portuguese - Tetun}   & BM25                                   & 0.4373         & 0.1282                & 0.3755                \\
                                   &                                       & DFR BM25                              & 0.4373         & 0.1283                & 0.3742                \\
                                   &                                       & Hiemstra LM                           & 0.4458         & 0.1335                & 0.3775                \\ \cdashline{2-6}
                                   & \multirow{3}{*}{English - Tetun}      & BM25                                   & 0.4508         & 0.1291                & 0.3696                \\
                                   &                                       & DFR BM25                              & 0.4492         & 0.1287                & 0.3684                \\
                                   &                                       & Hiemstra LM                           & 0.4542         & 0.1313                & 0.3738                \\ \cdashline{2-6}
                                   & \multirow{3}{*}{Indonesian - Tetun}   & BM25                                   & 0.4847         & 0.1406                & 0.4050                 \\ 
                                   &                                       & DFR BM25                              & 0.4847         & 0.1406                & 0.4050                 \\
                                   &                                       & Hiemstra LM                           & 0.4627         & 0.1323                & 0.3847                \\ \hline
\multirow{9}{*}{DeepSeek-Chat}     & \multirow{3}{*}{Portuguese - Tetun}   & BM25                                   & 0.4915         & 0.1509                & 0.4253                \\
                                   &                                       & DFR BM25                              & 0.4932         & 0.1513                & 0.4252                \\
                                   &                                       & Hiemstra LM                           & 0.5000         & 0.1527                & 0.4321                \\ \cdashline{2-6}
                                   & \multirow{3}{*}{English - Tetun}      & BM25                                   & 0.4746         & 0.1354                & 0.3921                \\
                                   &                                       & DFR BM25                              & 0.4729         & 0.1353                & 0.3911                \\
                                   &                                       & Hiemstra LM                           & 0.4576         & 0.1330                & 0.3850                 \\ \cdashline{2-6}
                                   & \multirow{3}{*}{Indonesian - Tetun}   & BM25                                   & 0.5237         & 0.1648                & 0.4454                \\
                                   &                                       & DFR BM25                              & 0.5237         & 0.1648                & 0.4443                \\
                                   &                                       & Hiemstra LM                           & 0.5220          & 0.1615                & 0.4376                \\ \hline
\multirow{9}{*}{Claude-3.7 Sonnet} & \multirow{3}{*}{Portuguese - Tetun}   & BM25                                   & 0.5661         & 0.1692                & 0.4987                \\
                                   &                                       & DFR BM25                              & 0.5695         & 0.1713                & 0.5011                \\
                                   &                                       & Hiemstra LM                           & \textbf{0.5797}         & \textbf{0.1787}                & \textbf{0.5091}                \\ \cdashline{2-6}
                                   & \multirow{3}{*}{English - Tetun}      & BM25                                   & 0.5576         & 0.1612                & 0.4702                \\
                                   &                                       & DFR BM25                              & 0.5593         & 0.1628                & 0.4700                  \\
                                   &                                       & Hiemstra LM                           & \textbf{0.5661}         & \textbf{0.1685}                & \textbf{0.4750}                 \\ \cdashline{2-6}
                                   & \multirow{3}{*}{Indonesian - Tetun}   & BM25                                   & 0.6034         & 0.1914                & 0.5168                \\ 
                                   &                                       & DFR BM25                              & \textbf{0.6051}        & \textbf{0.1922}                & \textbf{0.5174}                \\
                                   &                                       & Hiemstra LM                           & 0.5949         & 0.1859                & 0.5070                 \\ \hline
\multirow{9}{*}{LLaMA-3.3 70B}     & \multirow{3}{*}{Portuguese - Tetun}   & BM25                                   & 0.3729         & 0.1096                & 0.2940                 \\
                                   &                                       & DFR BM25                              & 0.3729         & 0.1103                & 0.2934                \\
                                   &                                       & Hiemstra LM                           & 0.3661         & 0.1068                & 0.2869                \\ \cdashline{2-6}
                                   & \multirow{3}{*}{English - Tetun}      & BM25                                   & 0.3271         & 0.0769                & 0.2623                \\
                                   &                                       & DFR BM25                              & 0.3271         & 0.0768                & 0.2615                \\
                                   &                                       & Hiemstra LM                           & 0.3254         & 0.0779                & 0.2643                \\ \cdashline{2-6}
                                   & \multirow{3}{*}{Indonesian - Tetun}   & BM25                                   & 0.3729         & 0.1096                & 0.2940                 \\
                                   &                                       & DFR BM25                              & 0.3729         & 0.1103                & 0.2934                \\
                                   &                                       & Hiemstra LM                           & 0.3661         & 0.1068                & 0.2869      \\\hline
\end{tabular}
}
\end{table}

Evaluating the retrieval performance for few-shot prompting with one example per language, Claude 3.7 Sonnet outperformed all other models across all languages and metrics, as shown in Table~\ref{tab:clir-tetun-few-shot}. Compared to the zero-shot results, providing one example per LLM call for Portuguese-to-Tetun translation improved retrieval performance by up to +0.93 percentage points. In contrast, for English-to-Tetun, the inclusion of an example resulted in a performance drop of up to -2.21 percentage points. For Indonesian-to-Tetun, the differences were marginal: a slight performance drop in P@10 and NDCG@10 (up to -0.63 percentage points), while MAP@10 showed a small gain of +0.29 percentage points.

\section{Discussion} \label{sec:discussion}
The effectiveness of CLIR in Tetun varies depending on the translation model and the prompting strategy applied to LLMs. In the zero-shot setting, Google Translate consistently demonstrated stronger retrieval performance overall, particularly for Indonesian-to-Tetun query translation. When paired with the Hiemstra LM, it achieved the highest scores among all zero-shot configurations evaluated. The effectiveness of Google Translate for query translation in CLIR for Tetun aligns with findings from the CLEF 2009 track~\cite{ferro-peters-clef-2009}, where its use substantially improved CLIR performance. Likewise, studies on English-to-Chinese CLIR reported retrieval results only marginally lower than monolingual ones when Google Translate was employed for query translation~\cite{wu-2010-query-translation-google-translate}.

Under the few-shot configuration, Claude 3.7 Sonnet showed the strongest retrieval performance for all languages, consistent with its strong zero-shot performance in Portuguese-to-Tetun translation, particularly at P@10 and MAP@10. For Portuguese-to-Tetun query translation, few-shot prompting led to a marginal improvement of up to +0.93 percentage points compared to the zero-shot setting. However, for English and Indonesian queries, even when the example was provided, LLM-based translations did not surpass the retrieval performance using Google Translate. These results underscore the effectiveness of Google Translate as a reliable translation tool for CLIR in Tetun under zero-shot settings.

When examining the relationship between translation quality and retrieval effectiveness, we observed findings consistent with those reported in the literature. For instance, although Google Translate achieved the highest BLEU score for Portuguese-to-Tetun translation (see Table~\ref{tab:bleu-scores}), it was outperformed by Claude 3.7 Sonnet in retrieval effectiveness, as shown in Table~\ref{tab:clir-tetun-zero-shot}. 
Similarly, while GPT-4.5-preview achieved the highest BLEU score for English-to-Tetun translation, it yielded lower retrieval performance compared to Google Translate. One possible explanation is that retrieval effectiveness depends not only on translation quality but also on how well the translated queries align with document representations and how effectively the retrieval and ranking models process those queries. Factors such as vocabulary overlap and model-specific ranking behavior may influence retrieval outcomes in ways that BLEU scores do not capture.
These findings address RQ1, suggesting that higher translation quality---as measured by BLEU---does not always lead to improved retrieval performance in CLIR. This observation aligns with prior studies by Pecina et al.~\cite{pecina-et-al-2014-mt-mlir-medical-domain} and Lignos et al.~\cite{lignos-et-al-challenge-mt-lrl}, both of which reported similar discrepancies between translation quality and retrieval performance across various domains and language pairs.

When comparing CLIR performance with the monolingual Tetun baselines, addressing RQ2, results remain significantly lower than those achieved by monolingual retrieval. This gap may be explained by the limitations in the translation models' capabilities. Notably, Google Translate only recently added support for Tetun, and the limited number of Tetun documents on the web as of 2023~\cite{de-jesus-nunes-2024-labadain30k-sigul,kudugunta-et-al-2023} further constrains the ability of LLMs to process the language effectively.

\textbf{Limitations and potential biases}: The volunteer translators are students, and not all have formal training in linguistics, which may affect the overall quality of the human translations. Additionally, no systematic validation of translation quality was conducted, potentially introducing errors or inconsistencies in the translated queries. Another limitation is the short length of the queries (between three and five words), which provides limited context and may hinder accurate translation, particularly in linguistically ambiguous cases. Furthermore, the translators’ backgrounds, language preferences, and interpretation styles may introduce bias into how certain terms or phrases are translated across languages.

\section{Conclusions and Future Work} \label{sec:conclusion}
This study explored the feasibility and effectiveness of CLIR for Tetun, an under-resourced language, by automatically translating queries from Portuguese, English, and Indonesian using both Google Translate and LLMs. The findings reveal several key insights.

First, despite advances in multilingual technologies, CLIR performance for Tetun remains significantly below that of monolingual retrieval. This highlights the ongoing challenges of supporting LRLs in information access tasks. Second, while LLMs show promise in handling a variety of tasks, including translation, their effectiveness is highly dependent on the source language and their familiarity with or knowledge of the target language. In some cases, traditional machine translation tools like Google Translate prove more reliable, particularly when newly supporting a language like Tetun.

Most critically, our study reveals that translation quality---as measured by standard metrics such as BLEU---does not correlate with retrieval effectiveness. This finding aligns with prior research~\cite{lignos-et-al-challenge-mt-lrl,pecina-et-al-2014-mt-mlir-medical-domain} and suggests the need to move beyond translation-centric metrics when evaluating CLIR systems. Retrieval success depends not only on linguistic translation accuracy but also on how well translated queries align with document representations and the retrieval models.

In future work, we aim to explore more effective CLIR systems for languages like Tetun by going beyond improvements in translation techniques. Our focus will include expanding Tetun’s digital footprint and rethinking retrieval strategies to better reflect its linguistic characteristics and resource constraints. We will also employ other translation evaluation metrics, such as ROUGE and METEOR, to more accurately assess translation quality and examine its correlation with retrieval effectiveness.

\begin{credits}
\subsubsection{\ackname} 
We thank Luisa Carlos, Zeferina da Gra\c{c}a, and Lara da Costa for translating the Tetun queries into English, Portuguese, and Indonesian, respectively.
This work is financed by national funds through FCT - Funda\c{c}\~{a}o para a Ci\^{e}ncia e a Tecnologia, under the support UID/50014/2023 (DOI: \href{https://doi.org/10.54499/UID/50014/2023} {10.54499/UID/50014/2023}). 
Gabriel de Jesus is supported by the Ph.D. studentship grant SFRH/BD/151437/2021, also from FCT (DOI: \href{https://doi.org/10.54499/SFRH/BD/151437/2021} {10.54499/SFRH/BD/151437/2021}).

\subsubsection{\discintname}
The authors have no competing interests to declare that are
relevant to the content of this article.
\end{credits}

%
%
\bibliographystyle{splncs04}
\bibliography{refs}

\end{document}